\documentclass{pasj01}

\usepackage{color}
\usepackage{natbib}

\Received{$\langle$reception date$\rangle$}
\Accepted{$\langle$acception date$\rangle$}
\Published{$\langle$publication date$\rangle$}

 \newcommand{\kms}{km s$^{-1}$\, } 
 \newcommand{\Hii}{H{$\,${\sc ii}}~} 

\begin{document}

\title{Interaction between Northern Coal Sack in the Cyg OB 7 cloud complex
and the multiple super nova remnants including HB 21}

\author{Kazuhito \textsc{Dobashi}\altaffilmark{1}, Tomomi \textsc{Shimoikura}\altaffilmark{1},
Nobuhiro \textsc{Endo}\altaffilmark{1}, Chisato \textsc{Takagi}\altaffilmark{1},
Fumitaka \textsc{Nakamura}\altaffilmark{2,3},
Yoshito \textsc{Shimajiri}\altaffilmark{4}, and
Jean-Philippe  \textsc{Bernard}\altaffilmark{5}}
\altaffiltext{1}{Department of Astronomy and Earth Sciences, Tokyo Gakugei University, Koganei, Tokyo  184-8501, Japan} 
\altaffiltext{2}{National Astronomical Observatory of Japan, Mitaka, Tokyo 181-8588, Japan}
\altaffiltext{3}{Department of Astronomical Science, School of Physical Science, SOKENDAI (The Graduate University for Advanced Studies), Osawa, Mitaka, Tokyo 181-8588, Japan}
\altaffiltext{4}{Laboratoire AIM, CEA/DSM-CNRS-Universit\'e Paris Diderot,
IRFU/Service d'Astrophysique, CEA Saclay, 91191, Gif-sur-Yvette, France}
\altaffiltext{5}{Universit\'e de Toulouse, Institut de Recherche en Astrophysique et Plan\'etologie, F-31028 Toulouse cedex 4, France}
\email{dobashi@u-gakugei.ac.jp}


\maketitle


\begin{abstract}
We report possible interaction between multiple super nova remnants (SNRs) and
Northern Coal Sack (NCS) which is a massive clump ($\sim1 \times 10^3$ $M_\odot$) in the Cyg OB 7 cloud complex
and is forming a massive Class 0 object.
We performed molecular observations of the $^{12}$CO$(J=1-0)$, $^{13}$CO$(J=1-0)$, and C$^{18}$O$(J=1-0)$
emission lines using the 45m telescope at the Nobeyama Radio Observatory, and we found that
there are mainly four velocity components at $V_{\rm LSR} \simeq -20$, $-6$, $-4$, and $10$ \kms.
The $-6$ and $-4$ \kms components correspond to the systemic velocities of NCS and
the Cygnus OB 7 complex, respectively, and the other velocity components originate from
distinct smaller clouds. Interestingly, there are apparent correlations and anti-correlations among the spatial distributions
of the four components, suggesting that they are physically interacting with one another.
On a larger scale, we find that a group of small clouds belonging to the $-20$ and $10$ \kms components
are located along two different arcs around some SNRs including HB 21 which has been suggested to be interacting
with the Cyg OB 7 cloud complex, and we also find that NCS is located right at the interface of the arcs.
The small clouds are likely to be the gas swept up by the stellar wind of the massive stars which created the SNRs.
We suggest that the small clouds alined along the two arcs recently encountered NCS and the massive star
formation in NCS was triggered by the strong interaction with the small clouds.

\end{abstract}


\section{Introduction}

Northern Coal Sack (NCS) is a well-known dark, opaque condensation in the Cygnus region.
Radio observations in 1990's have revealed that NCS is a massive clump and is forming a massive Class 0
object at the center; the Class 0 object is very bright in the millimeter/radio continuum \citep[e.g.,][]{Jenness1995,McCutcheon1991}, and is accompanied by an H$_2$O maser \citep{Miralles1994}
whose coordinates were found to be
$\alpha_{1950}={\rm 21^h07^m46.^s65}, \, \delta_{1950}={\rm +52^\circ 10^{'} 22.^{''}8}$
(corresponding to
$\alpha_{2000}={\rm 21^h09^m21.^s63}, \, \delta_{2000}={\rm +52^\circ 22^{'} 37.^{''}5}$, and
$l=92.^\circ6710, \, b=3.^\circ0714$)
by VLA measurements \citep{Jenness1995}.
Based on interferometric observations at millimeterwave, Bernard, Dobashi, and Momose (\citeyear{Bernard1999}) discovered that
the Class 0 object named GH2O 092.67+07 is associated with an extremely young and
compact molecular outflow with a dynamical
age of only $\sim3500$ years and a radius of $\sim8000$ AU.
They also found that its circumstellar disk around the Class 0 object
shows a clear collapsing motion with rotation.
In addition, there is another young massive star
creating a
compact \Hii region (IRAS 21078+5211) in NCS
in the vicinity of the Class 0 object.

NCS has a molecular mass of $\sim1 \times 10^3$ $M_\odot$ in total
which is a typical mass of clumps forming young clusters
\citep[e.g.,][]{Saito2007,Shimoikura2013}. 
According to our recent statistical studies of cluster forming clumps \citep{Shimoikura2016,Shimoikura2018},
the clump-scale collapsing motion with rotation as found toward NCS is commonly observed 
in an early stage of cluster formation. NCS is apparently a massive clump
just after the onset of such dynamical infall, and will eventually form a star cluster.

NCS is located in the direction of a giant molecular cloud (GMC) near the Cyg OB 7 association
\citep[e.g.,][]{Dame1985,Falgarone1987},
which we refer to as the Cyg OB 7 cloud complex, or more simply the Cyg OB 7 cloud in this paper.
Though their definite association has not been established well,
NCS is probably a part of the Cyg OB 7 cloud complex,
and we assume that NCS  is located at the same distance as the complex \citep[800 pc,][]{Humphreys1987}.

\begin{figure*}
\begin{center}
\includegraphics[scale=0.7]{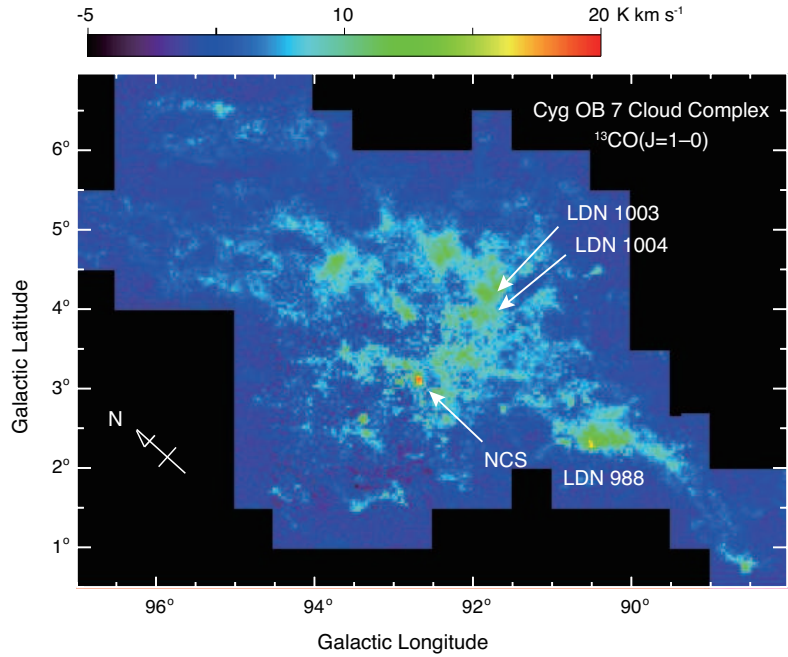}
\end{center}
\caption{
Integrated intensity map of the $^{13}$CO($J=1-0$) emission line revealing the entire extent
of the Cyg OB7 cloud complex. The map was obtained by the Nagoya 4m telescope \citep{Dobashi1994,Dobashi2014}.
Velocity range used for the integration is $-25<V_{\rm LSR}<15$ km s$^{-1}$.
Locations of NCS, LDN 988, LDN 1003, and LDN 1004 are indicated.
\label{fig:cygob7}}
\end{figure*}

In figure \ref{fig:cygob7}, we show the global distribution of the Cyg OB 7 cloud complex
revealed by large-scale $^{13}$CO($J=1-0$) observations \citep{Dobashi1994, Dobashi1996, Dobashi2014}.
The total mass of the Cyg OB 7 cloud complex derived from the $^{13}$CO data is $\sim1\times10^5$ $M_\odot$.
Other than NCS, there are a few known star forming sites in the complex
corresponding to opaque regions catalogued by \citet{Lynds1962}, e.g.,
LDN 988 \citep[e.g.,][]{Herbig2006, Movsessian2014}, LDN 1003 \citep[e.g.,][]{Aspin2009}, and
LDN 1004 \citep{Dobashi2014}, which are indicated in figure \ref{fig:cygob7}.
Compared to the Orion A cloud with a similar molecular mass \citep[e.g.,][]{Nagahama1998,Nishimura2015,Shimajiri2014,Shimajiri2015},
the Cyg OB 7 cloud is much less active in terms of star formation and is more turbulent \citep[e.g.,][]{Dobashi1994}.
As a GMC, the Cyg OB 7 cloud should be in an intermediate state between the Orion A and
the Maddalena cloud which is very turbulent and does not form young stars \citep[e.g.,][]{Maddalena1985}.
The Cygnus OB 7 cloud complex and NCS are therefore precious targets
to study how massive star formation is initiated in GMCs.

The purpose of the present study is to reveal the mass distributions and velocity field in and around
NCS, and to investigate what triggered the massive star formation in NCS.
It is known that the typical radial velocity of NCS \citep[$V_{\rm LSR}=-6$ km s$^{-1}$,][]{Bernard1999}
is slightly different from that of the rest of the Cyg OB 7 cloud \citep[$-4$ km s$^{-1}$,][]{Dobashi1994},
and it is also known that there are some small clouds at very different velocities
around NCS \citep{Dobashi1994}. 
These features in velocity may give us a hint to understand
the origin and nature of NCS.
It has been suggested by \citet{Tatematsu1990}
that the Cyg OB 7 cloud can be interacting with the super nova remnant (SNR) HB 21.
Their suggestion is very interesting, and the above features in velocities may have a relation with
the possible interaction.
However, a firm and direct observational evidence is needed to confirm the possible influence
of the SNR on the kinematics and star formation in NCS.

For the above purpose, we have carried out molecular observations with
the $^{12}$CO$(J=1-0)$, $^{13}$CO$(J=1-0)$, and C$^{18}$O$(J=1-0)$ emission lines
using the 45m telescope at the Nobeyama Radio Observatory (NRO). The observations were made
part of ``Star Formation Legacy Project" at the NRO (led by F. Nakamura) to observe
star forming regions such as
Orion A, Aquila Rift, and M17.
An overview of the project (Nakamura et al. 2018a, in preparation)
as well as detailed observational results for the individual regions
are given in other articles
(Orion A: Tanabe et al.  Ishii et al. 2018, in preparation, Nakamura et al. 2018b, in preparation, Takemura et al. 2018, in preparation,
Aquila Rift: Shimoikura et al. 2018b, in press, Kusune et al. 2018 in preparation,
M17: Shimoikura et al. 2018c, in preparation, Nugyen Luong et al. 2018, in preparation, and Sugitani et al. 2018, in preparation).

In this paper, we report results of the observations toward NCS.
we describe the observational procedures
in Section \ref{sec:observations}. We revealed the velocity field
around NCS, and also detected small clouds having distinct radial velocities.
Interestingly, we found clear correlations and anticorrelations in their spatial distributions.
We will present these results in Section \ref{sec:results}.
Based on the results, we discuss the origin of the small clouds as well as their influence on star formation
in NCS in Section \ref{sec:discussion}. The main conclusions of this paper are summarized in Section \ref{sec:conclusions}.


\section{Observations }\label{sec:observations}

We performed molecular observations toward NCS with the
$^{12}$CO($J=1-0$), $^{13}$CO($J=1-0$), and C$^{18}$O($J=1-0$) emission lines
using the 45 m telescope at the NRO
in the two periods from 2013 March to May and from 2016 February to May.
The receiver system named TZ \citep{Nakajima2013} was used to obtain the spectral data during the first period,
and another new receiver system named FOREST \citep{Minamidani2016} was used during the second period.
These receivers provided the total system noise temperature ($T_{\rm sys}$) typically in the range $150-400$ K
at 110 GHz depending on the molecular lines and the weather conditions.
As the backend, we used the digital spectrometers called SAM45 \citep{Kamazaki2012} having 4096 channels providing
a velocity resolution of $\sim 0.02$ km s$^{-1}$ at 110 GHz. 

Using the above set of the systems, we performed mapping observations covering an area of
$\sim 8\arcmin \times 8\arcmin$ around NCS along equatorial coordinates.
The mapping observations were made using the
On-The-Fly \citep[OTF,][]{Sawada2008} technique. The obtained raw data were processed
in a standard way using the software package NOSTAR available at the NRO.
We resampled the data onto a $10\arcsec$ grid
using a Gaussian convolution function
to produce the spectral data cube having an angular resolution of $\sim23\arcsec$ (FWHM)
and velocity resolution of 0.05 km s$^{-1}$.

For the intensity calibration, we used the standard chopper-wheel method \citep{Kutner1981}
to scale the data to units of $T_{\rm a}^*$, and then further scaled the data to $T_{\rm mb}$
by applying the beam efficiency of the 45m telescope at 110-115 GHz.
The stability of the system was checked to be accurate within 10\%
by observing a small region around GH2O 092.67+07 everyday. 
The pointing accuracy is better than $5\arcsec$ as was checked by observing
the SiO maser T-Cep every 2 hours during the observations.

The noise level of the resultant spectral data is about
$\Delta T_{\rm rms}=1.3$, $0.9$, and $0.9$ K for the $^{12}$CO, $^{13}$CO, and C$^{18}$O
emission lines, respectively, in units of $T_{\rm mb}$ for the velocity resoluton of 0.05 km s$^{-1}$.


\section{Results}\label{sec:results}

\subsection{Molecular distributions and velocity components}
\begin{figure*}
\begin{center}
\includegraphics[scale=0.7]{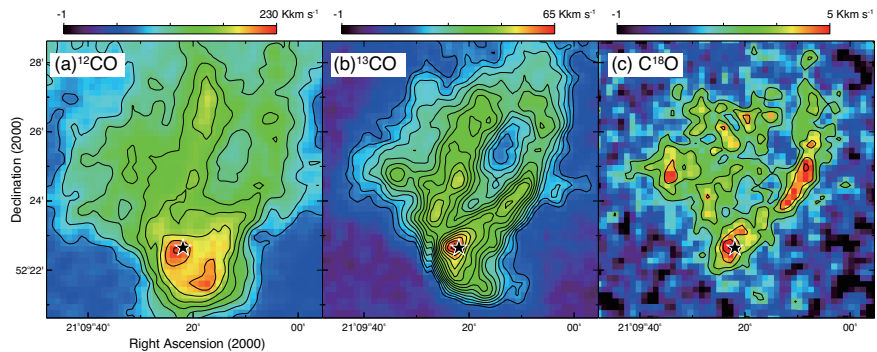}
\end{center}
\caption{
Integrated intensity maps of the (a) $^{12}$CO$(J=1-0)$, (b) $^{13}$CO$(J=1-0)$, and (c) C$^{18}$O$(J=1-0)$
emission lines around NCS observed with the 45m telescope.
Velocity ranges used for the integrations are
$-25<V_{\rm LSR}<15$ km s$^{-1}$,
$-11<V_{\rm LSR}<4$ km s$^{-1}$, and
$-10<V_{\rm LSR}<-4$ km s$^{-1}$
for the $^{12}$CO, $^{13}$CO, and C$^{18}$O emission lines, respectively.
Star symbol denotes the position of the massive Class 0 object GH2O 092.67+03.07.
Contours start from
80, 20, and 2 K kms$^{-1}$
with an increment of 
20, 3, and 1 K kms$^{-1}$
in panels (a), (b), and (c), respectively.
\label{fig:45mmaps}}
\end{figure*}

We show the obtained intensity maps of the $^{12}$CO, $^{13}$CO, and C$^{18}$O
emission lines in figure \ref{fig:45mmaps}.
As seen in the figure, the distribution of the optically thick $^{12}$CO  emission line shows a round, cometary
shaped structure of NCS, whereas those of the optically thinner $^{13}$CO and C$^{18}$O emission lines
exhibit more filamentary structures. The signal-to-noise ratio
is rather poor in the C$^{18}$O map, but two main filaments extending to the northwest
can clearly be seen in the $^{13}$CO map.

\begin{figure*}
\begin{center}
\end{center}
\includegraphics[scale=0.7]{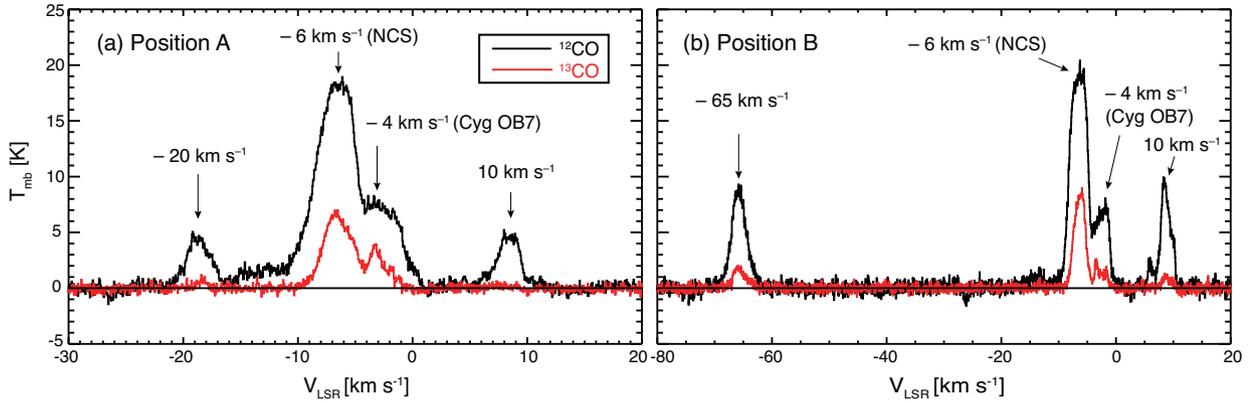}
\caption{
The $^{12}$CO (black lines) and $^{13}$CO (red lines) spectra observed at the positions A and B indicated
in figures \ref{fig:components}c and \ref{fig:components}d.
In the region observed with the 45m telescope, we find five velocity
components at $V_{\rm LSR}\simeq-65$, $-20$, $-10$, $-6$, $-4$ and $10$ km s$^{-1}$.
The emission lines at $V_{\rm LSR}\simeq-6$ km s$^{-1}$ and $-4$ km s$^{-1}$ originate from
the main body of NCS and the Cygnus OB cloud complex, respectively.
\label{fig:spe}}
\end{figure*}

In the observed region, we detected five velocity components in the $^{12}$CO and $^{13}$CO
emission lines. We will refer to them as $-65$, $-20$, $-6$, $-4$, and $10$ km s$^{-1}$ components
according to their typical LSR velocities ($V_{\rm LSR}$) in this paper.
Figure \ref{fig:spe} displays two examples of the observed spectra.
According to the previous studies of the Cyg OB 7 cloud \citep{Dobashi1994,Dobashi2014}
and NCS \citep{ Bernard1999},
the $-4$ km s$^{-1}$ and $-6$ km s$^{-1}$  components correspond to the characteristic velocities
of the entire Cyg OB 7 cloud complex and the main body of NCS, respectively.
The difference between the $-6$ km s$^{-1}$ and  $-4$ km s$^{-1}$ components is
rather small, and they are partially merged to each other as seen in the figure,
suggesting a close connection between these two components.
On the other hand, the $-65$, $-20$, and $10$ km s$^{-1}$ components are well separated in velocity.
At a glance, these velocity components appear as if they are coming from more distant clouds
unrelated to the Cyg OB 7 cloud complex, because the observed region is at low galactic latitudes ($b\simeq3^\circ$)
and can be contaminated by many other clouds lying on the same line-of-sight.
However, as we will show in the next subsection, two of them ($-20$, and $10$ km s$^{-1}$ components)
are apparently interacting with NCS and the Cyg OB 7 cloud.

We show the spatial distributions of the five velocity components in figure \ref{fig:components}.
In the figure, the $-65$, $-6$, and $-4$ km s$^{-1}$ components are shown by the $^{13}$CO intensity,
and the $-20$ and $10$ km s$^{-1}$ components are shown by the $^{12}$CO intensity
because they are week in $^{13}$CO.

Under the assumption of Local Thermodynamic Equilibrium (LTE),
we estimated the molecular mass of each velocity component
from the $^{12}$CO and $^{13}$CO data in a standard
way \citep[e.g.,][]{Shimoikura2011,Shimoikura2018}:
We first fitted the observed $^{12}$CO and $^{13}$CO spectra with
a single gaussian function, and determined the excitation temperature
of the molecules $T_{\rm ex}$ from the fitted brightness temperature
of the $^{12}$CO spectra assuming that the line is optically thick.
We then estimated the optical depth of the $^{13}$CO emission line
from the derived $T_{\rm ex}$ and the fitted gaussian parameters
of the $^{13}$CO spectra to derive $N$($^{13}$CO) the column density of $^{13}$CO.
We converted $N$($^{13}$CO) to
$N$(H$_2$) the column density of molecular hydrogen using the empirical relation
$N$(H$_2$)$=5\times10^5$$N$($^{13}$CO) \citep{Dickman1978}, and calculated the mass of the velocity
components assuming that all of them are located at a distance of 800 pc.
We repeated the above process pixel by pixel for every velocity component
over the mapped region. For pixels where $T_{\rm ex}$ estimated from the $^{12}$CO spectra
is lower than 10 K, we assumed a flat excitation temperature of $T_{\rm ex}=10$ K because
the optically thick assumption for the $^{12}$CO line is not valid for such pixels.
We note that the $^{13}$CO emission line for the $-20$ km s$^{-1}$ component is too weak
to perform a reliable mass estimate. For this velocity component, we therefore estimated
$N$(H$_2$) (and then the mass) from the $^{12}$CO intensity as $N$(H$_2$)$=X_{\rm CO}W_{\rm CO}$
where $W_{\rm CO}$ is the velocity-integrated intensity of the $^{12}$CO line in units of K km s$^{-1}$ and
$X_{\rm CO}$ is a conversion factor taken to be $1.8\times10^{20}$ cm$^{-2}$K$^{-1}$km$^{-1}$s \citep{Dame2001}.

We summarize the results of the above mass determination in table \ref{tab:mass}.
Within the area shown in figure \ref{fig:components}, NCS has a mass of $\sim1 \times 10^3$ $M_\odot$
which is a typical mass of dense massive clumps forming young clusters
\citep[e.g.,][]{Saito2007,Shimoikura2013,Shimoikura2018}. 
Except for the $-4$ km s$^{-1}$ component originating from the Cyg OB 7 cloud,
the other components at high velocities ($-65$, $-20$, and $10$ km s$^{-1}$) have
much smaller masses of an order of $\sim10$ $M_\odot$ if they are located at the same distance
as the Cyg OB 7 cloud complex.
In the table, we also list the maximum values of $T_{\rm ex}$ in each velocity component.
A high value of $T_{\rm ex}$($\simeq 50$ K) is found for the $-6$ km s$^{-1}$ component,
which is due to the massive Class 0 object forming at the center of NCS and
is consistent with the previous measurement \citep{Bernard1999}.

\subsection{Spatial correlations}
Except for the $-65$ km s$^{-1}$ component,
we found that there are apparent spatial correlations and anticorrelations among the identified velocity components.
As can be seen in figure \ref{fig:components},
the anticorrelations are especially obvious.
In the four panels (a)--(d) of figure \ref{fig:correlation},
we indicate some of the correlations and anticorrelations we found
by circles with white and pink broken lines, respectively, and describe them in the following:

\begin{enumerate}

\item {\bf The 10 and -6 km s$^{-1}$ components (figure \ref{fig:correlation}a)} \\
The most intense bump of the $-6$ km s$^{-1}$ component (red contours)
coincides with a valley of the $10$ km s$^{-1}$ component (color scale)
as denoted by the pink circle. In addition, one of the filaments of the $-6$ km s$^{-1}$ component
extending to the northwest coincides with a ridge of the $10$ km s$^{-1}$ component
around the white circle.

\item {\bf The 10 and the -20 km s$^{-1}$ components (figure \ref{fig:correlation}b)} \\
The $-20$ km s$^{-1}$ component (white contours) is detected in the northern part of the
mapped region, and it clearly coincides with a hole of the 10 km s$^{-1}$ component (color scale).

\item {\bf The -4 and 10 km s$^{-1}$ components (figure \ref{fig:correlation}c)} \\
The $10$ km s$^{-1}$ component (orange contours) sits in a large valley of the $-4$ km s$^{-1}$
component (color scale). The valley may also corresponds to the intense region of the $-6$ km s$^{-1}$
component (red contours in panels a and d).

\item {\bf The -6 and -20 km s$^{-1}$ components (figure \ref{fig:correlation}d)} \\
The distributions of the $-6$ and $-20$ km s$^{-1}$ components partially overlap,
and a ridge of the $-6$ component (red contours) matches with a valley of the
$-20$ km s$^{-1}$ component (white contours).

\end{enumerate}

A clear anticorrelation of different velocity components was first detected in
the Sgr B2 star forming cloud complex by 
\citet{Hasegawa1994}, which they interpreted as an evidence of
a cloud-cloud collision that induced the active star formation in the complex.
Since then, cloud-cloud collisions have been found to be a common phenomenon
which seems to significantly influence on star formation in molecular clouds
\citep[e.g.,][]{Torii2011,Fukui2018,Nishimura2018}.

Cloud-cloud collisions are often recognized through an anticorrelation between intensity
distributions of two velocity components, because one cloud penetrates
the other making a set of a ``bump" and ``hole" in their intensity maps.
In addition to such anticorrelations, we propose that correlations like the one
indicated in figure \ref{fig:correlation}a (by the circle with white broken line)
should be an evidence of cloud-cloud collisions as well, because
a diffuse cloud (e.g., an extended H{$\,${\sc i}} cloud with no or little CO)
colliding with a compact molecular cloud
should strip molecular gas from the compact cloud
when they pass through each other, which we should see as correlation
in the CO intensity maps. We suggest that the correlation
in figure \ref{fig:correlation}a was created in this manner.

Numerical simulations of cloud-cloud collisions predict that
there must be gas having intermediate velocities of the natal clouds
around the interface
\citep[e.g., see figures 3 and 5 of][illustrating a head-on collision of two clouds]{Habe1992}. 
We looked for such gas with intermediate velocities in the $^{12}$CO spectra
around where we can recognize the correlations and anticorrelations, and found
that there is $^{12}$CO emission with intermediate velocity between
the $-20$ and $-6$ km s$^{-1}$ components.
We display the spectrum in figure \ref{fig:spe},
and show the intensity distribution
of the intermediate velocity component in figure \ref{fig:inter_co}a.
We also show a position-velocity (PV) diagram in figure \ref{fig:inter_co}b
taken across the intersection of the $-20$ and $-6$ km s$^{-1}$ components.
The $^{12}$CO emission with intermediate velocities bridging the two velocity components
is obvious in the PV diagram.
The existence of the intermediate velocity component gives a support to our hypothesis
of cloud-cloud collisions at least between the $-20$ and $-6$ km s$^{-1}$ components.
However, we note that the intersection of the two components is the only place
where we can see CO emission with this intermediate velocity.
This may imply that the gas at intermediate velocity is not able to emit the CO lines in general,
because the CO molecules are dissociated by the collision or cannot be excited due to the low
gas densities.

To summarize, we found correlations and anticorrelations among the spatial distributions
of the $-20$, $-6$, $-4$, and $10$ km s$^{-1}$ components, which indicate that the four velocity
components are interacting. We further found CO emission with intermediate velocities
located at the intersection of the $-20$ and $-6$ km$^{-1}$ spatial distribution, which supports
the idea of interactions between the two components.

On the other hand, the $-65$ km s$^{-1}$ component is rather isolated showing no clear correlations or anticorrelations
with other components. We conclude that the $-65$ km s$^{-1}$ component likely originates from a distant
cloud unrelated to the Cyg OB 7 cloud complex.


\section{Discussion}\label{sec:discussion}

Among the four velocity components showing the correlations and anticorrelations,
the $-6$ km s$^{-1}$and $-4$ km s$^{-1}$ components are known to
originate from the massive NCS ($\sim 1\times 10^3$ $M_\odot$)
and the Cyg OB 7 cloud ($\sim 1 \times 10^5$ $M_\odot$), and thus the results in section \ref{sec:results}
indicate that much smaller clouds at $-20$ km s$^{-1}$ and $10$ km s$^{-1}$
are colliding against
NCS and the Cyg OB 7 cloud. In the following, we investigate the nature and origin of these
velocity components as well as their influence on star formation in NCS and the Cyg OB 7 cloud.

Because the area observed with the 45m telescope is quite limited,
we first investigate the distributions of the $-20$ km s$^{-1}$ and $10$ km s$^{-1}$ components
on a larger scale using the $^{13}$CO data obtained with the Nagoya 4m telescope \citep{Dobashi1994,Dobashi2014}.
Figure \ref{fig:snr} displays the locations of these components superposed on the intensity maps of
the $-4$ km s$^{-1}$ component tracing the Cyg OB 7 cloud and
the $-6$ km s$^{-1}$ component including NCS.
For comparison, we show the locations of the SNRs catalogued by \citet{Green2014}.

As can be seen in the figure, there are a number of small clouds belonging to
the $-20$ km s$^{-1}$ and $10$ km s$^{-1}$ components, and it is striking that
they are apparently alined along the arcs indicated by the orange and white broken lines, respectively.
One of the arcs (for the $-20$ km s$^{-1}$ component) is centered on
HB 21 which has a diameter and a dynamical age of $\sim2^\circ$ \citep{Green2014}
and $\sim5 \times 10^3$ yr \citep{Lazendic2006}, respectively.
This SNR has been suggested to be interacting with the Cyg OB 7 cloud \citep{ Tatematsu1990}.
Around the center of the other arc, there are also some SNRs
such as DA 551, 3C434.1, and G096.0+02.0.

If the small clouds are distributed randomly regardless of the SNRs in the area shown in figure \ref{fig:snr},
a probability for their alining on the arcs centered on the SNRs by chance would be very small. We therefore
interpret the coincidence between the arcs and the SNRs location as an evidence for physical interaction.
The small clouds along the arcs, however, are likely not influenced directly by the blast wave of the SNRs,
because the sizes of the SNRs measured in the radio continuum \citep{Green2014} are much smaller than the radii of the arcs,
and the small clouds along the arcs are well separated from the extents of the SNRs which are expressed by
the while circles in figure \ref{fig:snr}.
We believe that the small clouds are the gas swept up and/or accelerated by the stellar wind of the massive stars
which ultimately produced the SNRs. Note that a dynamical time scale of $\sim4$ Myr is obtained if we divide
the apparent radii of the arcs ($\sim 60$ pc) by the velocity difference between the arcs and
the Cygnus OB 7 cloud ($\sim15$ km s$^{-1}$), which is compatible with a lifetime of O type stars.

In addition, it is interesting to note that NCS is located at the interface of the two arcs.
This suggests a likely scenario that the small clouds belonging to both of the $-20$ km s$^{-1}$ and 10 km s$^{-1}$
components recently
encountered NCS, which should compress NCS sufficiently and
trigger the gravitational collapse of the clump and then star formation therein.
Such compression of molecular clouds by multiple shells or bubbles may be common
on a galactic scale, and it must give a significant effect on star formation
\citep[e.g.,][]{Inutsuka2015}.

The above scenario is favorable to find a solution to the following two problems.
One is the problem of ``colliding filaments" in LDN 1004.
In our earlier study, we identified a clump of $\sim1\times 10^4$ $M_\odot$
in LDN 1004 based on the near-infrared extinction maps \citep{Dobashi2011,Dobashi2013},
and we named it ``L1004E" \citep{Dobashi2014,Matsumoto2015}.
Molecular observations with the C$^{18}$O emission line revealed that
the clump contains several massive filaments of a few times $100$ $M_\odot$ each,
and that the filaments are colliding against one another.
However, according to MHD simulations we performed,
massive filaments can form spontaneously in a collapsing clump,
but collisions of the filaments scarcely happen,
unless they are forced to collide by strong compression from the outside \citep[see discussion of][]{Dobashi2014}.
The question is: what is the compressing force?
The clump L1004E is located just inside of the arc for the $-20$ km s$^{-1}$ component (see figure \ref{fig:snr}b),
and the arc must have crossed L1004E recently, which is a likely answer to the problem of
the unknown compressing force.

The other problem is the trigger for the onset of gravitational collapse of massive clumps to form star clusters.
In our recent statistical study of cluster formation based on a sample of more than 20 massive clumps
\citep{Shimoikura2018},
we found that, in terms of chemical composition, the clumps forming only a few stars which we classified as ``Type 1" clumps
are significantly older than the other clumps already forming clusters which we classified as ``Type 2" and ``Type 3" clumps
(see figures 13 and 14 of the paper).
At a glance, this finding is opposite to our expectation; we would expect that Type 1 clumps
should be the youngest among the Type 1 -- Type 3 clumps and will eventually form clusters
to turn into Type 2 clumps. We believe that the puzzling chemical compositions of the Type 1 clumps
mean that most of the Type 1 clumps are actually old being gravitationally stable for
a long time (due to cloud-supporting force, e.g., by the magnetic field) and they can start collapsing when
the equilibrium is broken by chance, e.g., by a sudden increment of the external pressure
due to the compression by SNRs or stellar wind from massive stars.
In other words, such external triggers may be needed for contraction of massive clumps to form clusters.
However, it should be very difficult to find a massive clump just after the onset of contraction
because of their very short free-fall time ($\sim 10^5$ yr).
NCS should be one of such rare examples, and we suggest that this kind of interactions
between massive clumps and SNRs and/or stellar wind of massive stars 
may be one of the major sources to trigger the gravitational collapse of massive clumps to form
star clusters.


\section{Conclusions} \label{sec:conclusions}

We performed molecular observations of Northern Coal Sack (NCS) in the Cyg OB 7 cloud complex
using the 45m telescope at the Nobeyama Radio Observatory (NRO),
and we investigated the global molecular distributions
and velocity field of NCS. The main conclusions of this paper can be summarized in the following points.

\begin{figure*}
\begin{center}
\includegraphics[scale=0.7]{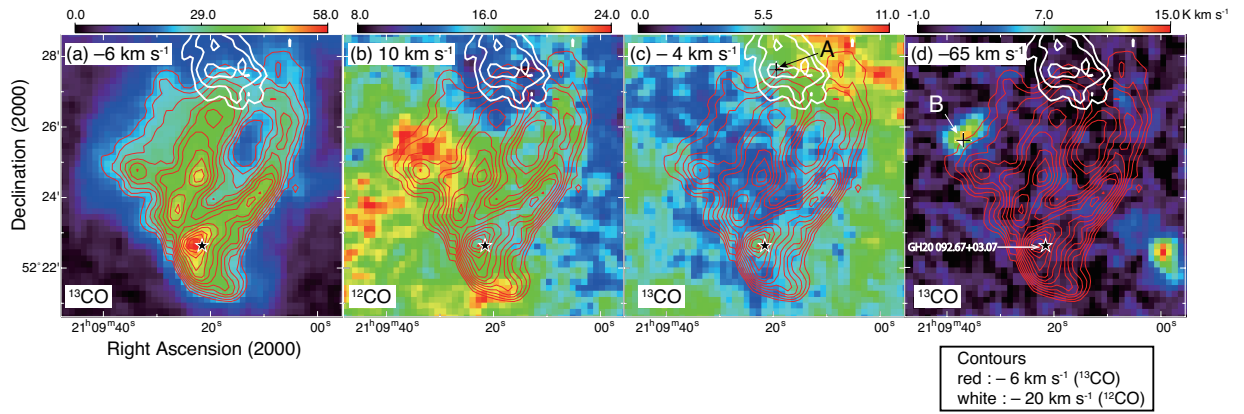}
\end{center}
\caption{
Distributions of the five velocity components identified in this study.
Color scales in panels (a), (b), (c), and (d) show intensities of the
$-6$, $10$, $-4$, and $-65$ km s$^{-1}$ components,
which are integrated over the velocity ranges
$-11<V_{\rm LSR}<-4$ km s$^{-1}$,
$7<V_{\rm LSR}<10$ km s$^{-1}$,
$-4<V_{\rm LSR}<1$ km s$^{-1}$, and
$-69<V_{\rm LSR}<-62$ km s$^{-1}$, respectively.
Common white contours are for the $-20$ km s$^{-1}$ component
integrated over the velocity range
$-20<V_{\rm LSR}<-17$ km s$^{-1}$,
and are drawn from
9 K km s$^{-1}$ with a step of 3 K km s$^{-1}$.
Red contours in panel (a), drawn from 21 K km s$^{-1}$ with a step of 3 K km s$^{-1}$,
are also shown in all of the panels for comparison.
The $^{12}$CO data are used to create the maps for the $-20$ and $10$ km s$^{-1}$ components,
and the $^{13}$CO data are used for the other components.
\label{fig:components}}
\end{figure*}

\begin{figure*}
\begin{center}
\includegraphics[scale=0.7]{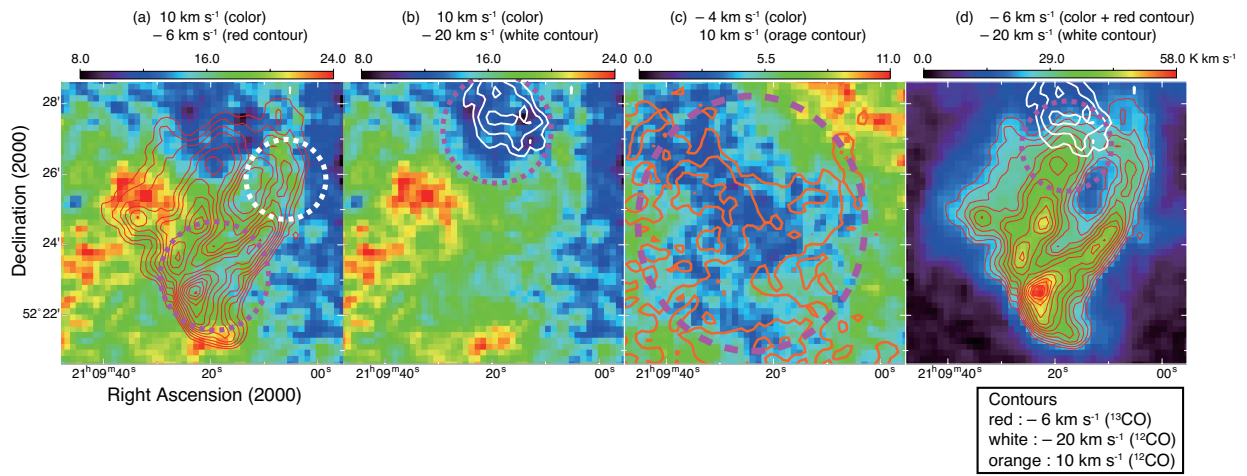}
\end{center}
\caption{
Circles with thick pink broken line indicate apparent anticorrelations between
(a) the 10 and $-6$ km s$^{-1}$ components,
(b) the 10 and $-20$ km s$^{-1}$ components,
(c) the $-4$ and $10$ km s$^{-1}$ components, and 
(d) the $-6$ and $-20$ km s$^{-1}$ components.
Circles with thick white broken line in panel (a) indicate a correlation between the 10 and $-6$ km s$^{-1}$ components. 
Red and white contours are the same as those in figure \ref{fig:components}.
Thick orange contours in panel (c) represent the $^{12}$CO intensity for the 10 km s$^{-1}$
component integrated over
the velocity range $7<V_{\rm LSR}<10$ km s$^{-1}$, and are drawn from 1 K km s$^{-1}$
with a step of 5 K km s$^{-1}$.
\label{fig:correlation}}
\end{figure*}

\begin{figure}
\begin{center}
\includegraphics[scale=.9]{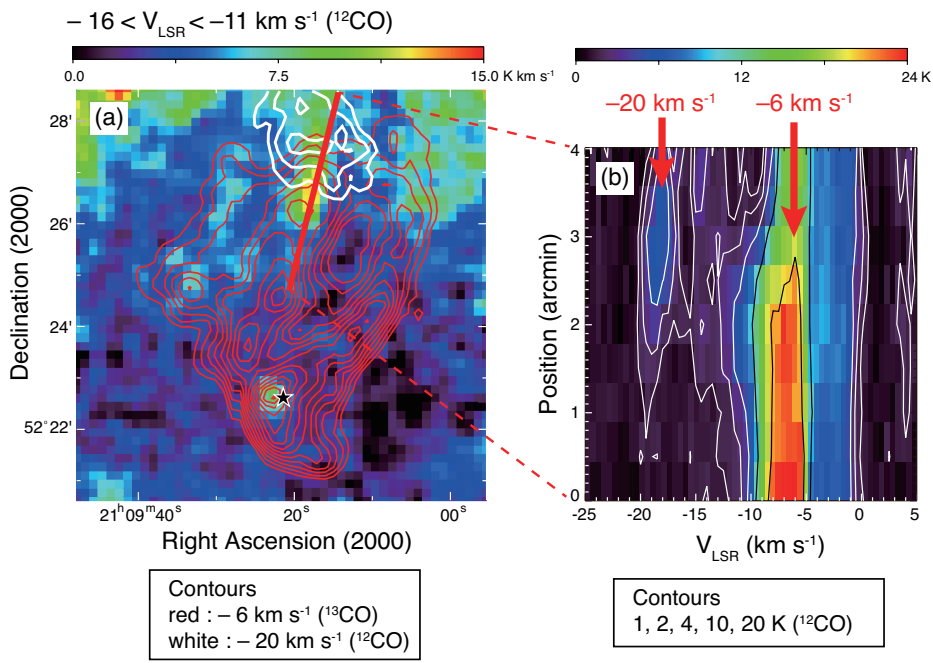}
\end{center}
\caption{
(a) Intensity distribution of the $^{12}$CO emission
integrated over $-16<V_{\rm LSR}<-11$ km s$^{-1}$.
The emission in this velocity range is evident around the position
where the $-6$ km s$^{-1}$ component (red contours) and $-20$ km s$^{-1}$
component (white contours) show an anticorrelation.
The red and white contours are the same as in figure \ref{fig:components}.
(b) Position-velocity diagram of the $^{12}$CO emission taken along the red cut
in panel (a). Contour are drawn at 1, 2, 4, 10, and 20 K.
Faint $^{12}$CO emission bridging the $-20$ and $-6$ km s$^{-1}$ components
is seen.
\label{fig:inter_co}}
\end{figure}

\begin{figure*}
\begin{center}
\includegraphics[scale=.7]{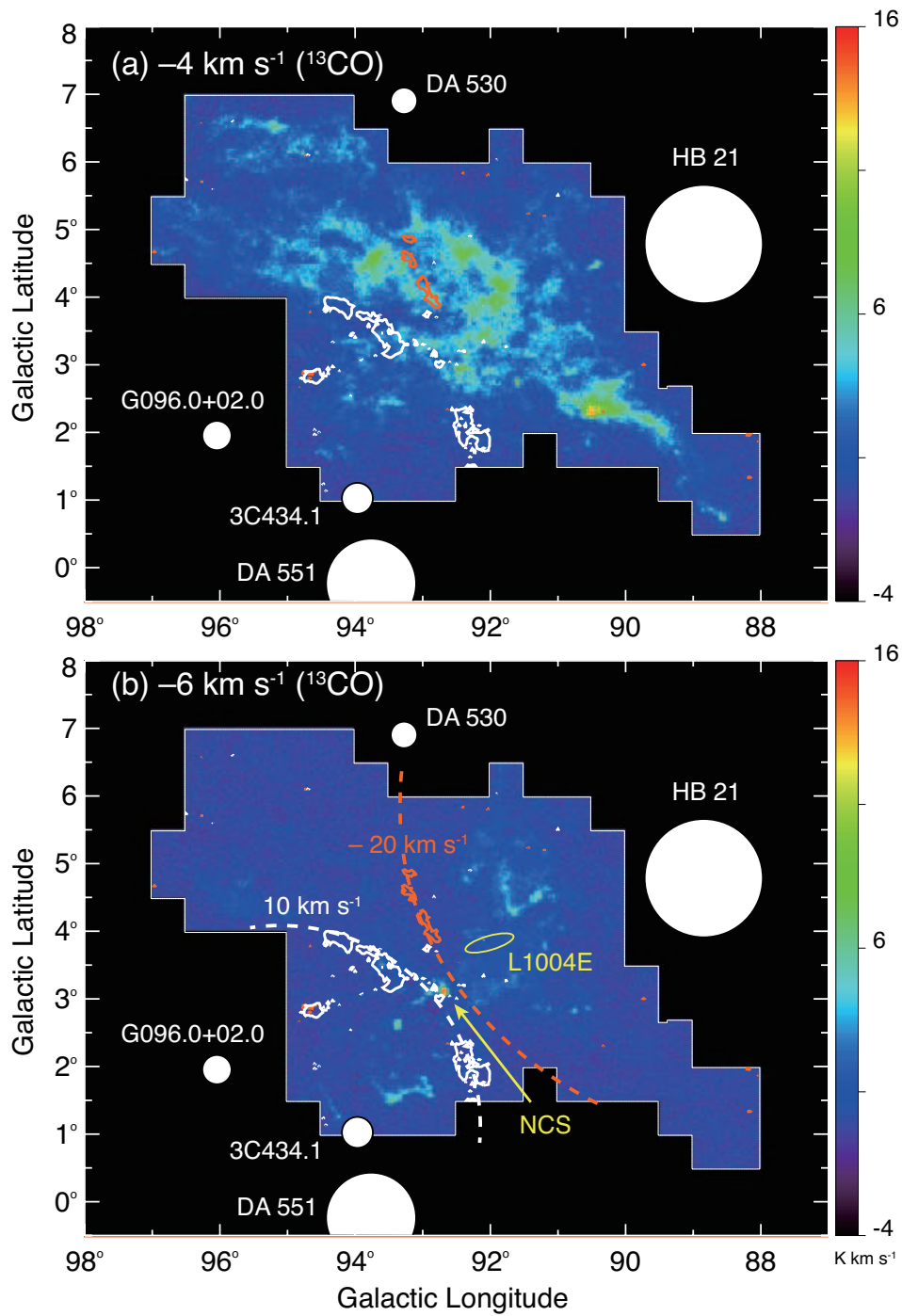}
\end{center}
\caption{
Distributions of the SNRs \citep[][white circles]{Green2014},
the $10$ km s$^{-1}$ component (white contours), and
the $-20$ km s$^{-1}$ component (orange contours)
shown on the large-scale $^{13}$CO maps of
(a) the $-4$ km s$^{-1}$ component tracing the Cyg OB 7 cloud complex and
(b) the $-6$ km s$^{-1}$ component tracing NCS.
Diameters of the white circles denote the catalogued sizes of the SNRs.
The white and orange contours start from 1 K km s$^{-1}$ with a step of 5 K km s$^{-1}$.
Velocity ranges used for the integration for the
$-20$, $-6$, $-4$, and $10$ km s$^{-1}$ components are the same as those in
figure \ref{fig:components}.
As shown in panel (b), the the $-20$ and $10$ km s$^{-1}$ components are lying along
the arcs denoted by the orange and white broken lines, and NCS is located at
their interface.
A yellow ellipse denotes the location of another dense clump in the Cyg OB 7 cloud complex (L1004E)
which is inferred to be influenced by an external compressing force  \citep{Dobashi2014}.
\label{fig:snr}}
\end{figure*}

\begin{enumerate}

\item We mapped an area of $\sim 8\arcmin \times 8\arcmin$ around NCS
with the $^{12}$CO($J=1-0$), $^{13}$CO($J=1-0$), and C$^{18}$O($J=1-0$) emission lines.
Based on the $^{12}$CO and $^{13}$CO data, we estimated the total molecular mass of NCS to be $\sim1 \times 10^3$ $M_\odot$
within the mapped region shown in figure \ref{fig:components}.

\item We found that there are five velocity components in the mapped region, which we named $-65$, $-20$,
$-6$, $-4$, and $10$ km s$^{-1}$ components after their typical LSR velocities.
The $-6$ and $-4$ km s$^{-1}$ components correspond to the main body of NCS and the Cyg OB 7 cloud complex, respectively.
The $-20$ and $10$ km s$^{-1}$ components trace small clouds around NCS, and we found that
there are apparent correlations and anticorrelations among the distributions of the $-20$, $-6$,  $-4$, and $10$ km s$^{-1}$
components, indicating that these clouds are interacting with one another.
The origin of the $-65$ km s$^{-1}$ component is unknown, and it may originate from a distant cloud unrelated to the Cyg OB 7 cloud complex.

\item We investigated the distributions of the $-20$ and $-4$ km s$^{-1}$ components on a large scale, and found that
clouds belonging to the velocity components are aligned along two arcs. Interestingly, around the center of the arcs, there are
super nova remnants (SNRs) including HB 21 which has been suggested to be interacting with the Cyg OB 7 cloud complex.
We found that NCS is located right at the intersection of the arcs. We suggest the $-20$ and $10$ km s$^{-1}$ components
are tracing small clouds swept up by the stellar wind of massive stars which created the SNRs and that they are
currently crossing the NCS region, producing strong compression and triggering gravitational collapse and subsequent
high mass star formation.

\end{enumerate}


\begin{ack}
This work was financially supported by JSPS KAKENHI Grant Numbers
JP17H02863, JP17H01118, JP26287030, and JP17K00963. 
The 45 m radio telescope is operated by NRO, a branch of NAOJ. 
YS received support from the ANR (project NIKA2SKY, grant agreement ANR-15-CE31-0017).
\end{ack}




\clearpage


\begin{table}
\caption{Mass of the Velocity Components} 
\begin{tabular}{ccr}  
\hline\noalign{\vskip3pt} 
	\multicolumn{1}{c}{Components} & \multicolumn{1}{c}{Max. $T_{\rm ex}$} & \multicolumn{1}{c}{Mass} \\
	\multicolumn{1}{c}{} & \multicolumn{1}{c}{(K)} & \multicolumn{1}{c}{($M_\odot$)}  \\  [2pt] 
\hline\noalign{\vskip3pt} 
$-65$ km s$^{-1}$	&	$24$	&	$15$ 	\\
$-20$ km s$^{-1}$	&	$10$	&	$17$ 	\\
$-6$ km s$^{-1}$	&	$52$	&	$1022$ 	\\
$-4$ km s$^{-1}$	&	$16$	&	$182$ 	\\
$10$ km s$^{-1}$	&	$16$	&	$34$ 	\\
\hline
\end{tabular} \label{tab:mass}
   \begin{tabnote}
The table lists the mass of the five components within the region displayed in figure \ref{fig:components}.
The mass of the $-20$ km s$^{-1}$ component was calculated from the intensity of the $^{12}$CO emission line,
and those of the other components were calculated from the column density of $^{13}$CO (see text). 
The second column lists the maximum value of the excitation temperature of each component
estimated from the $^{12}$CO emission line.
   \end{tabnote}
\end{table} 
\clearpage




\end{document}